 \documentclass[smallabstract,smallcaptions]{dccpaper}

\usepackage{epsfig}
\usepackage{citesort}
\usepackage{amsmath}
\usepackage{amssymb}
\usepackage{color}
\usepackage{url}

\newlength{\figurewidth}
\newlength{\smallfigurewidth}

\begin{document}

\title
{\large
\textbf{Optimizing Audio Compression Through Entropy-Controlled Dithering}
}

\author{%
Ellison Murray, Morriel Kasher, and Predrag Spasojevic\\[0.5em]
{\small
\begin{minipage}{\linewidth}
\centering
\begin{tabular}{c}
Rutgers University\\
New Brunswick, NJ, USA\\
\url{eem144@scarletmail.rutgers.edu}\\
\url{mbk94@scarletmail.rutgers.edu}\\
\url{spasojev@winlab.rutgers.edu}
\end{tabular}
\end{minipage}
}
}

\maketitle
\thispagestyle{empty}

\
\begin{abstract}
This paper explores entropy-controlled dithering techniques in audio compression, examining the application of both standard and modified TPDFs, combined with noise shaping and entropy-controlled parameters, across various audio contexts, including variations in pitch, loudness, rhythm, and instrumentation. Perceptual quality metrics such as VISQOL and STOI were used to evaluate performance. The results demonstrate that TPDF-based dithering consistently outperforms RPDF, particularly under optimal alpha conditions, while highlighting performance variability based on signal characteristics. These findings suggest the situational appropriateness of using various TPDF distributions. This work emphasizes the trade-off between entropy and perceptual fidelity, offering insights into the potential of entropy-controlled dithering as a foundation for enhanced audio compression algorithms. A practical implementation as a Digital Audio Workstation plugin introduces customizable dithering controls, laying the groundwork for future advancements in audio compression algorithms.
\end{abstract}

\Section{Introduction}

Achieving exceptional audio quality is a central pursuit across diverse communities, including audiophiles, musicians, film enthusiasts, and fans eager to experience their favorite artists' latest releases in pristine fidelity. As a musician working extensively with digital audio workstations, I have continually sought methods to optimize output quality while managing storage constraints—a challenge that resonates with many professionals in the field.\

\indent Currently, lossless audio compression formats such as Apple Lossless Audio Codec (ALAC) and FLAC deliver high-quality sound~\cite{Ahmed2018}.  However, these formats demand significantly more storage than their lossy counterparts, posing practical limitations for applications like streaming and large-scale storage~\cite{Ng1998}. With the growing demand for premium audio quality, balancing quality and storage efficiency has become increasingly critical.\

\indent Recent research has explored distortion-controlled dithering in image compression, focusing on optimizing the trade-off between signal distortion and perceptual quality. This work introduced various distortion measures, including MSE, MACE\textsuperscript{2}, and MSCE, to establish a framework for designing distortion-controlled dithered quantization systems, demonstrating its application in uniform dithering for image compression~\cite{Kasher2024}.\

\indent While several studies have investigated dithered quantization in audio and speech applications~\cite{Floros2006},~\cite{Borsky2016}, they fall short of addressing optimal dithering strategies for compression and distortion. These studies often overlook the contexts where dithering is most appropriate, leaving a gap in understanding its full potential in audio compression.\

\indent This paper explores entropy-controlled dithering in audio compression, specifically employing Triangular Probability Density Functions (TPDFs) over Uniform or Rectangular Probability Density Functions (RPDFs). TPDFs are preferred for their ability to effectively decouple noise, reducing the perceptual prominence of artifacts compared to RPDFs~\cite{Vanderkooy1987}. This paper aims to achieve near-lossless audio compression by leveraging this technique while preserving perceptual quality.\

\indent The scope of this research extends beyond evaluating compressibility and quality trade-offs in processed audio files. It seeks to assess the contexts in which this approach is most effective, considering additional techniques such as noise shaping. This paper aims to advance audio compression methods by addressing these objectives, making high-quality audio more accessible without compromising fidelity.\

\Section{Background}
Quantization is a process used to compress data files by reducing the storage space required~\cite{Gray1998}. However, this process comes with a significant drawback: the loss of data. In certain types of compression, such as audio, this loss also leads to a decline in perceptual quality. For this reason, quantization is classified as a lossy compression method. This paper focuses on perceptual quality loss during quantization, which manifests as artifacts such as ringing, rattling, graininess, and other unwanted distortion~\cite{Menkman2011}, which are direct consequences of data loss during quantization.\ 

\indent The mean squared error (MSE) is considered for this project; however, it is not the primary focus of optimization since it does not account for perceptual distortions, but rather measures the similarity between the processed and distorted signals \cite{Borsky2016}. MSE is calculated using Equation 1, where $x$ is the original audio signal, and $\hat{x}$ is the processed audio signal. While this metric is useful for evaluating how much of the original audio information is retained, it does not fully capture perceptual audio quality \cite{Herre1999}. In practical applications, minimizing this error corresponds to maximizing the preservation of the original audio signal's fidelity to its intended form. \begin{equation}
MSE = d(x, \hat{x}) = (x - \hat{x})^2
\end{equation} 

\begin{figure}[h]
\centering
\includegraphics[width=4in]{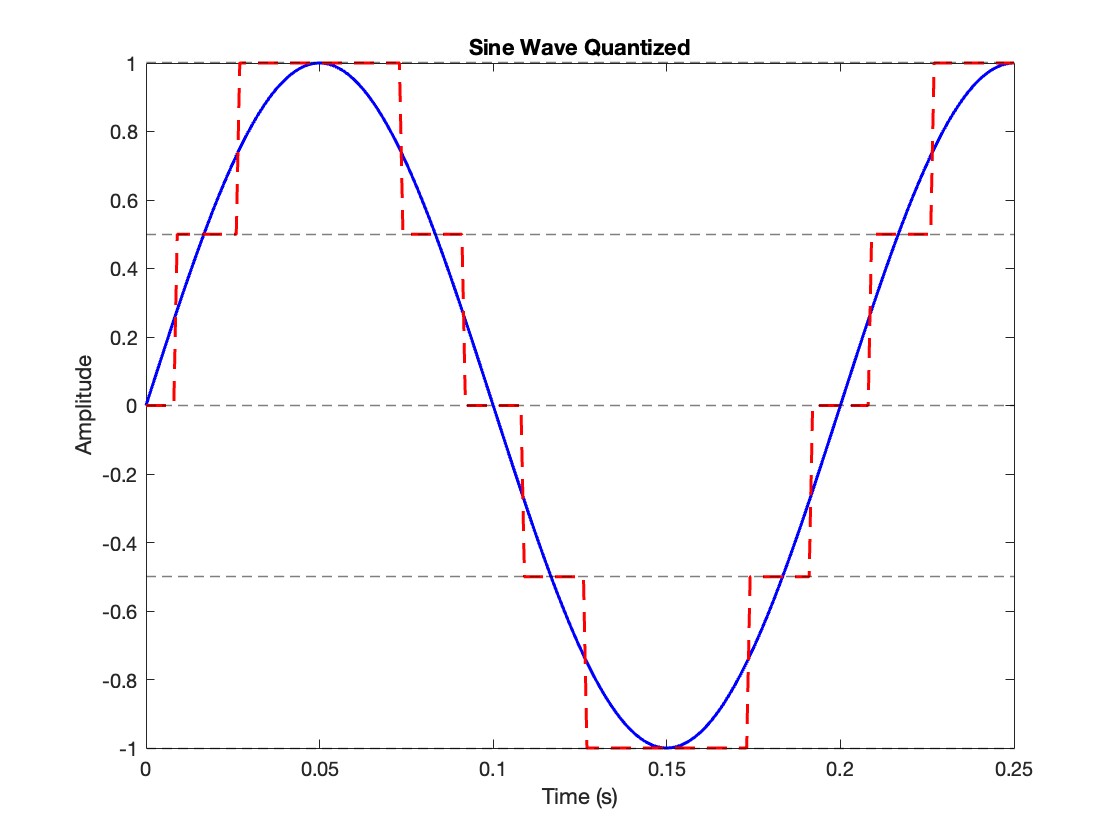}
\caption{\label{fig:1}%
Inputted sine wave with amplitude 1 (blue) overlaid by quantized sine wave to 2 bits (red).}
\end{figure}

\indent With the quantization process, the size to which the audio file is truncated is controlled by the number of bits that the audio signal is quantized. The bits are what determines the number of quantization levels, as it’s a function of the bits, given $2^b$. So, for 8-bit quantization, the number of quantization levels will be 256. The way that the truncation in quantization works is by approximating the continuous signal amplitude values to the discrete quantization level values (See Fig.~\ref{fig:1}). The quantization employed in this paper involved reducing audio files, originally stored as 8-bit integers, to 3-bit resolution, resulting in 8 quantization levels.\ 

\indent A technique called dithering is often employed to mitigate the loss of perceptual quality introduced through quantization. Dithering involves adding a small, random noise signal to the data before quantization~\cite{Gray1998}. This additional noise smooths the transitions between quantization levels, effectively reducing perceptual artifacts and making them less noticeable. The comparison below highlights the differences between dithered quantization and regular quantization.\ 

\indent Although dither signals are perceptible, they are generally not considered disruptive or "annoying" to listeners. However, they can still slightly affect perceptual quality. Noise shaping can be applied to further optimize the dithering process. Noise shaping shifts the dither signal into frequency ranges where it is less perceptible to the human ear, based on the equal-loudness contour, which reflects the human ear's sensitivity to different frequencies~\cite{Herre1999}. This refinement enhances the effectiveness of dithering while maintaining high perceptual quality.\ 

\begin{figure}[h]
\centering
\includegraphics[width=4in]{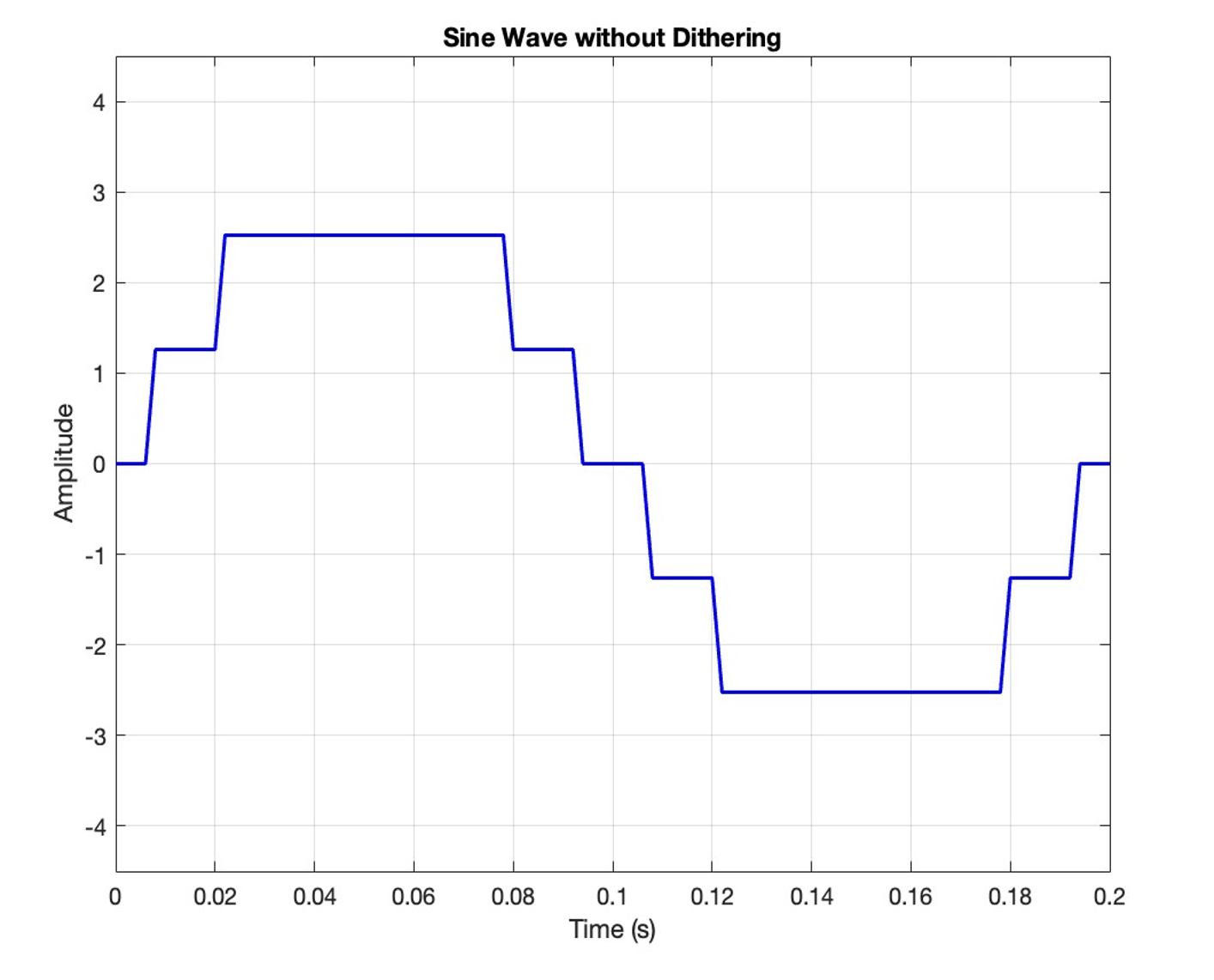}
\caption{\label{fig:2}%
Quantized sinusoidal wave without dither applied.}
\end{figure}

\indent Without dithering, quantization becomes rough, distorting the original shape and resulting in a loss of detail. This quantization error manifests as perceptual artifacts due to its high correlation with the audio signal (See Fig.~\ref{fig:2}).\

\begin{figure}[h]
\centering
\includegraphics[width=4in]{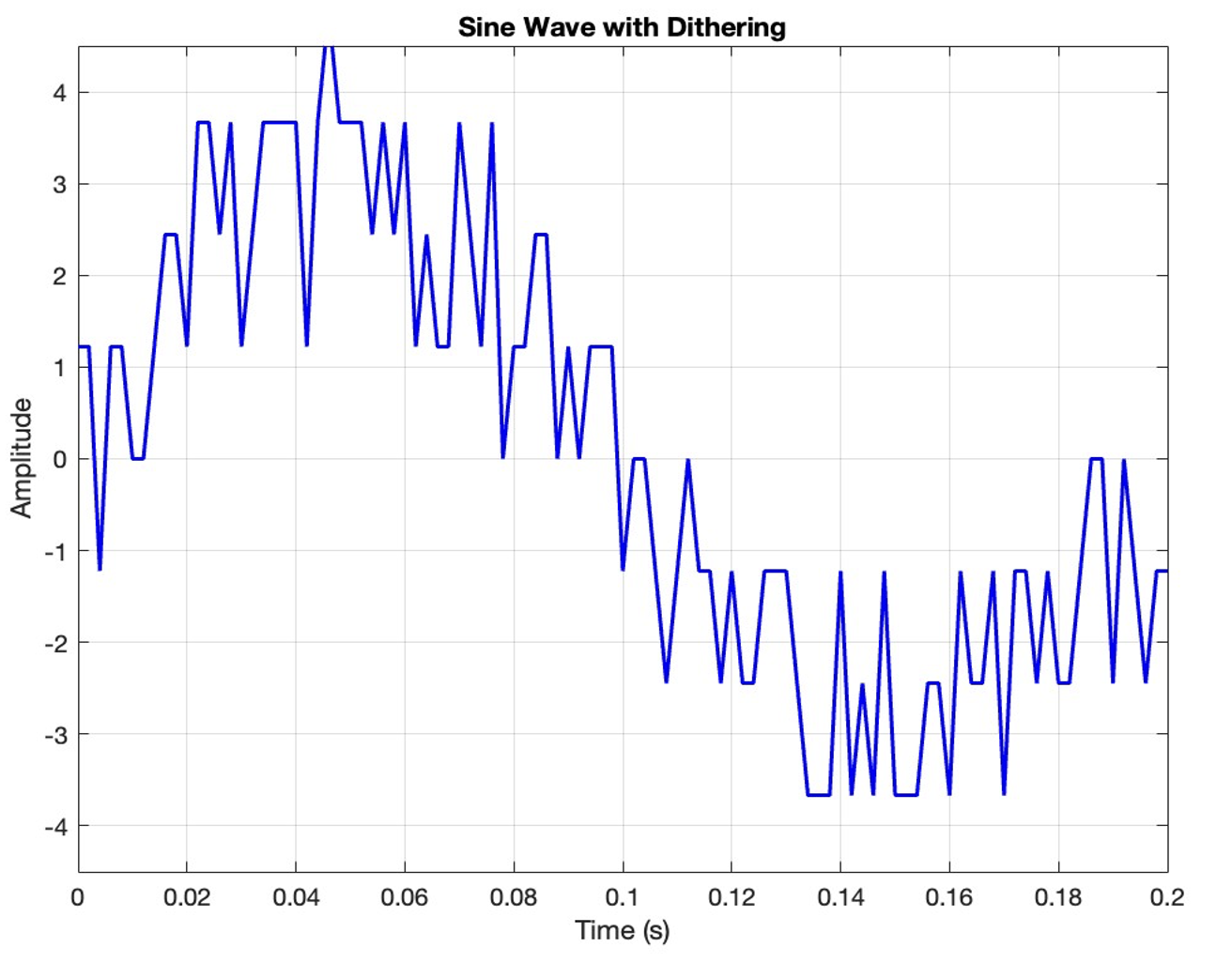}
\caption{\label{fig:3}%
Quantized sinusoidal wave with dither applied.}
\end{figure}

\indent Applying dithering retains more of the sine wave's original shape by preserving the statistical behavior of the noise. (See Fig.~\ref{fig:3}) This process not only breaks the correlation between the signal and quantization error but also converts distortion into broadband noise instead of structured artifacts. Consequently, the quantization error becomes less perceptible to the human ear or measurement systems~\cite{Gray1998}.\ 

\indent On the other hand, data compression relies on identifying and exploiting redundancies within the data to reduce its size. When dithering is applied, the increased randomness within the signal reduces these redundancies, making the data harder to compress~\cite{Kulkarni2014}.\ 

\indent Balancing this trade-off between perceptual quality and compressibility requires careful consideration. The application of dithering introduces more randomness—or entropy—into the signal, directly impacting its perceived quality and compressibility. Equation 2 is used to calculate entropy in bits. \begin{equation}
H(x) = -\sum_{i=1}^M p_i \log_b p_i
\end{equation}
\indent The trade that is subject to this project’s exploration can be modeled through the relationship described in Equation 3, where there is a parameter that controls the size of the dither signal, $f_v$ is the dither signal as a function of $\alpha$, and $P(f_v)$ and $C(f_v)$ are representative of functions for the perceptual quality and compressibility respectively.\begin{equation}
(1-\alpha)P(f_v) + \alpha C(f_v)
\end{equation}

\begin{figure}[h]
\centering
\includegraphics[width=4in]{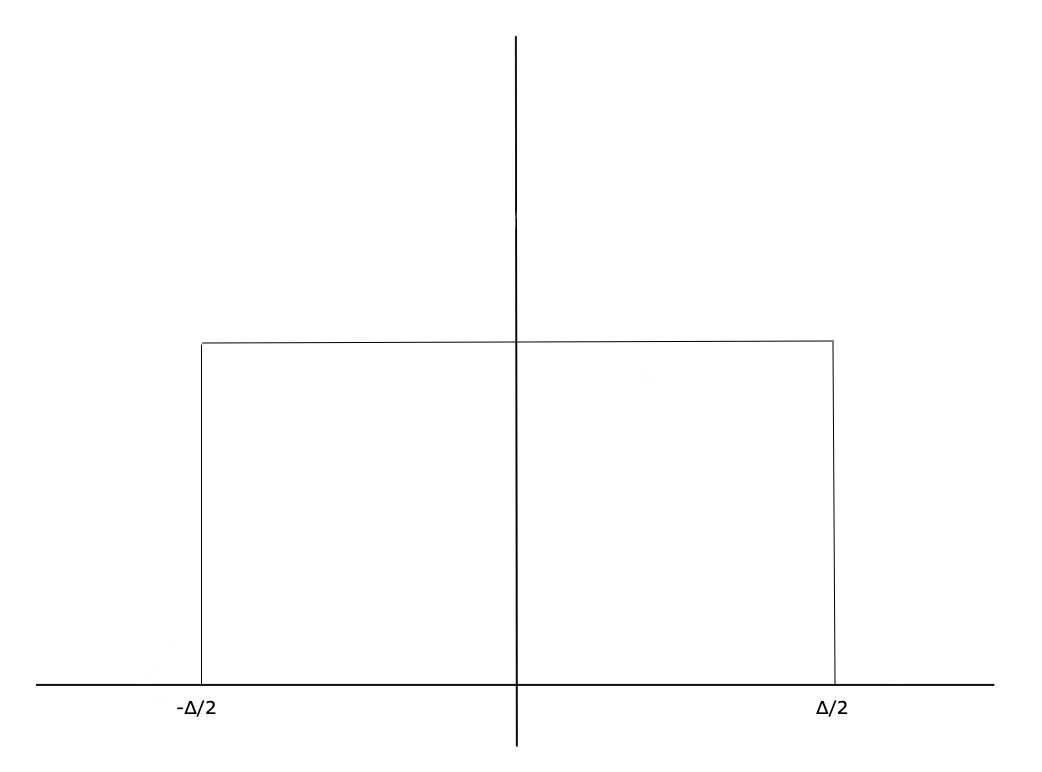}
\caption{\label{fig:4}%
Rectangular Probability Density Function.}
\end{figure}

\indent Dither signals can follow various probability density functions (PDFs), such as Gaussian, rectangular, and triangular distributions, to name a few. The PDF of a signal describes the likelihood of the signal's amplitude taking on different values. For example, in a uniform rectangular PDF $f_v(v) = \Pi_{\Delta}(v)$, the dither signal's amplitude ranges from $-\Delta/2$ to $+\Delta/2$, and the probability of each amplitude within that range is equal. This contrasts with a triangular PDF, where the probabilities are not uniform (see Fig.~\ref{fig:4} and Fig. 5). 

\begin{figure}[t]
\centering
\begin{tabular}{cc}
\includegraphics[width=2.7in]{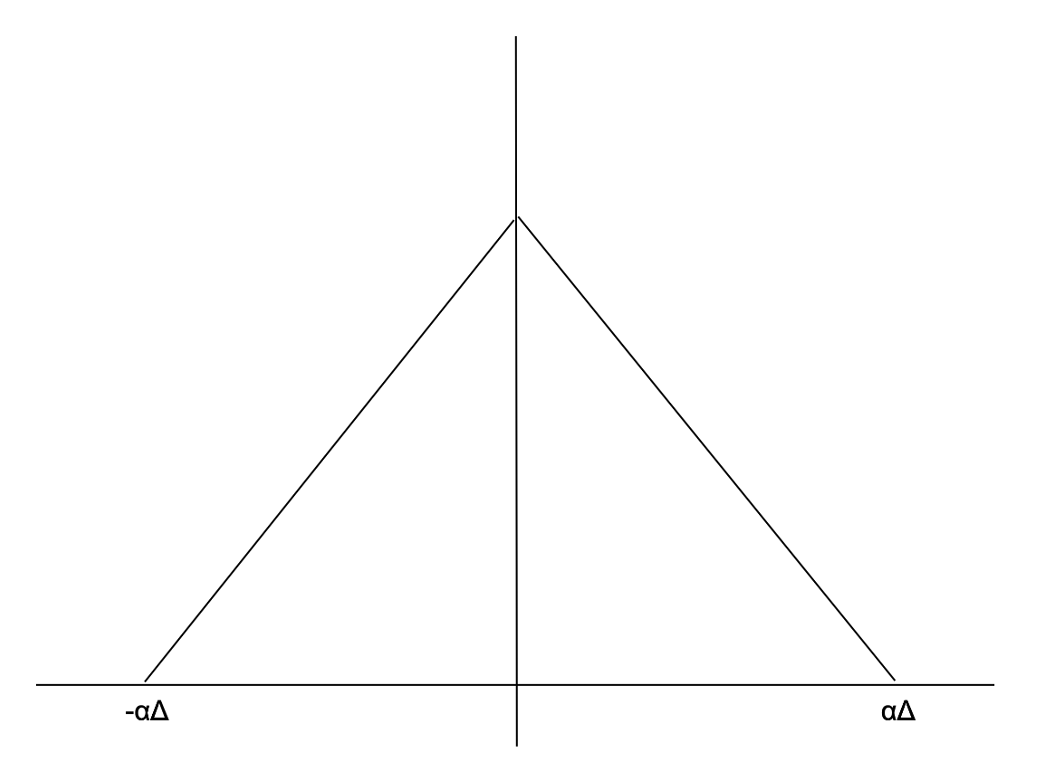} &
\includegraphics[width=2.7in]{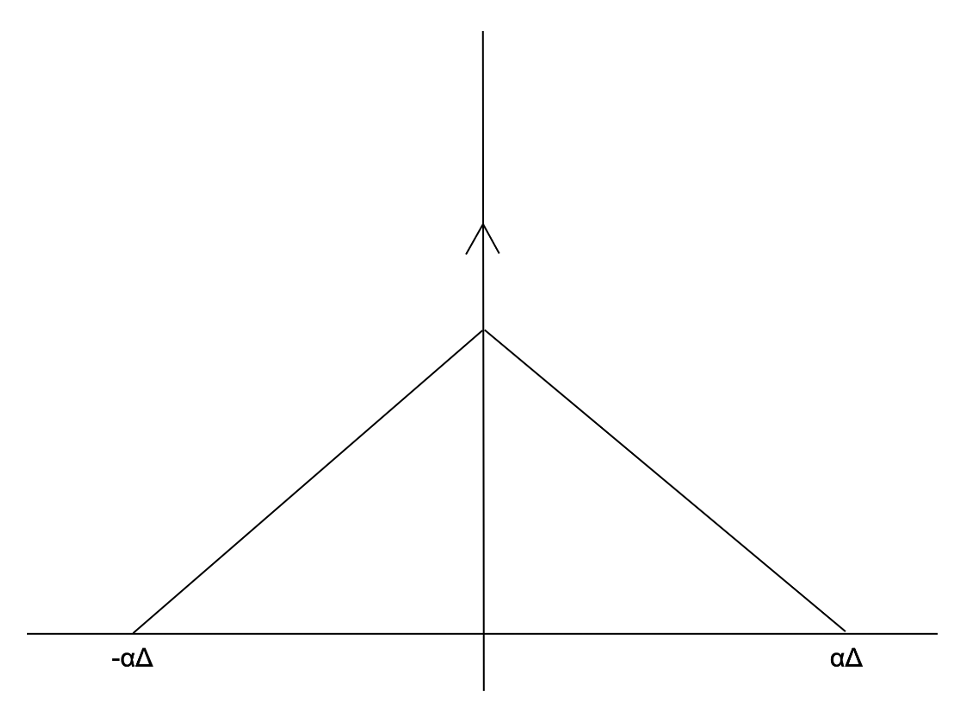} \\
{\small (a)} & {\small (b)}
\end{tabular}
\caption{\label{fig:5}%
(a) Triangular Probability Density Function (b) Modified Triangular Probability Density Function.}
\end{figure}

\Section{Method}
The two dither distributions tested were a standard TPDF signal, $f_v = \Lambda_{2\alpha\Delta}(v)$, and a modified TPDF signal (see Fig.~\ref{fig:5}), both distributed uniformly across the frequency spectrum. The standard TPDF signal has an amplitude ranging from $-\alpha\Delta$ to $+\alpha\Delta$, in which $\Delta$ represents the quantization step size. The step size is determined using Equation 4, based on the amplitude of the input audio signal and the number of quantization bits.\begin{equation}
\Delta = 2A/2^b
\end{equation}
\indent The TPDF dither is generated by computing the discrete differences of an input RPDF dither. Initialization begins by setting the first TPDF element equal to the first RPDF value, with subsequent elements calculated as the differences between consecutive RPDF values. This process simulates a convolution, effectively transforming RPDF into TPDF.\

\indent The modified TPDF is defined as $f_v = \alpha\Lambda_{2\alpha\Delta}(v) + (1 - \alpha) \delta(v)$, where the generated TPDF dither, weighted by $\alpha$, is combined with a Dirac delta function at the center of the quantization level, weighted by $(1 - \alpha)$ (see Fig. 5). This introduces a discrete component to an otherwise continuous PDF, which is expected to increase redundancies within the signal.\ 

\SubSection{Setup}
All signal processing for the project was conducted in MATLAB, with testing divided into four distinct parts. The first part focused on testing the technique at varying loudness levels, using a sinusoidal wave of C4 with loudness ranging from -25 dB to 0 dB. The sinusoidal signals tested in this section were generated in Logic Pro using the Test Oscillator and Quick Sampler plugins, recorded at 24 bits, and exported as 8-bit stereo, interleaved .wav files sampled at 44.1 kHz, with no dithering added. When exported from Logic Pro, samples were not normalized. The second part investigated sinusoidal wave pitches ranging from C4 to C5 (frequencies 261.63 Hz to 523.25 Hz), incremented by half steps and tested individually.\ 

\indent The third section analyzed audio files containing chords, which featured multiple pitches played simultaneously. This included testing timbre using sinusoidal waves, trumpets, and pianos samples. Each sample comprised a C major triad with octave doubling (C4, E4, G4, C5). The fourth section focused on samples with rhythmic elements, incorporating bass riffs, piano segments, drum loops, and a speech sample extracted from a podcast to ensure high-quality audio. Stock loops from Logic Pro were utilized, including City Nights Trumpet Lead, Disco Breaking Drums, 12 Bar Blues Bass, and 70s Rock Piano 01. For all tests except those related to loudness, the audio samples were exported at -10 dB. Loudness adjustments were managed in Logic Pro, which measures decibels (dB) relative to a 0 dB-FS reference level, the maximum digital signal level before clipping occurs.\ 

\begin{figure}[h]
\centering
\begin{tabular}{cc}
\includegraphics[width=2.8in]{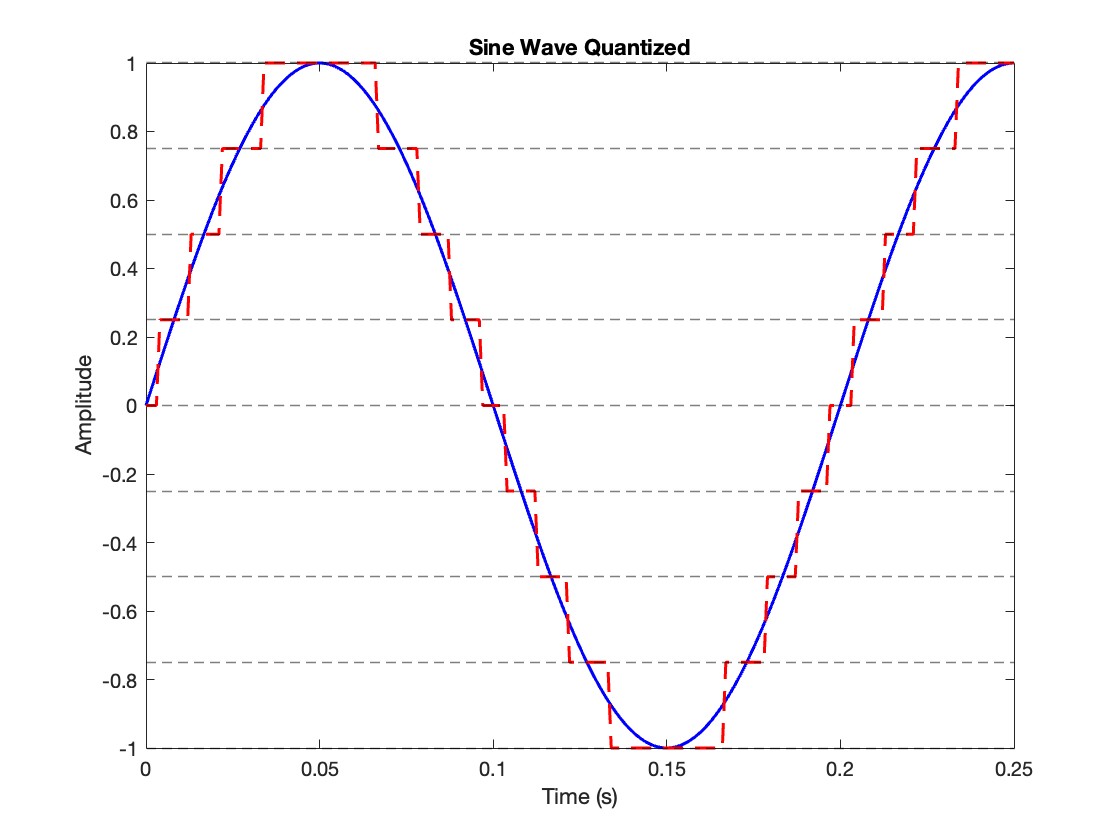} &
\includegraphics[width=2.8in]{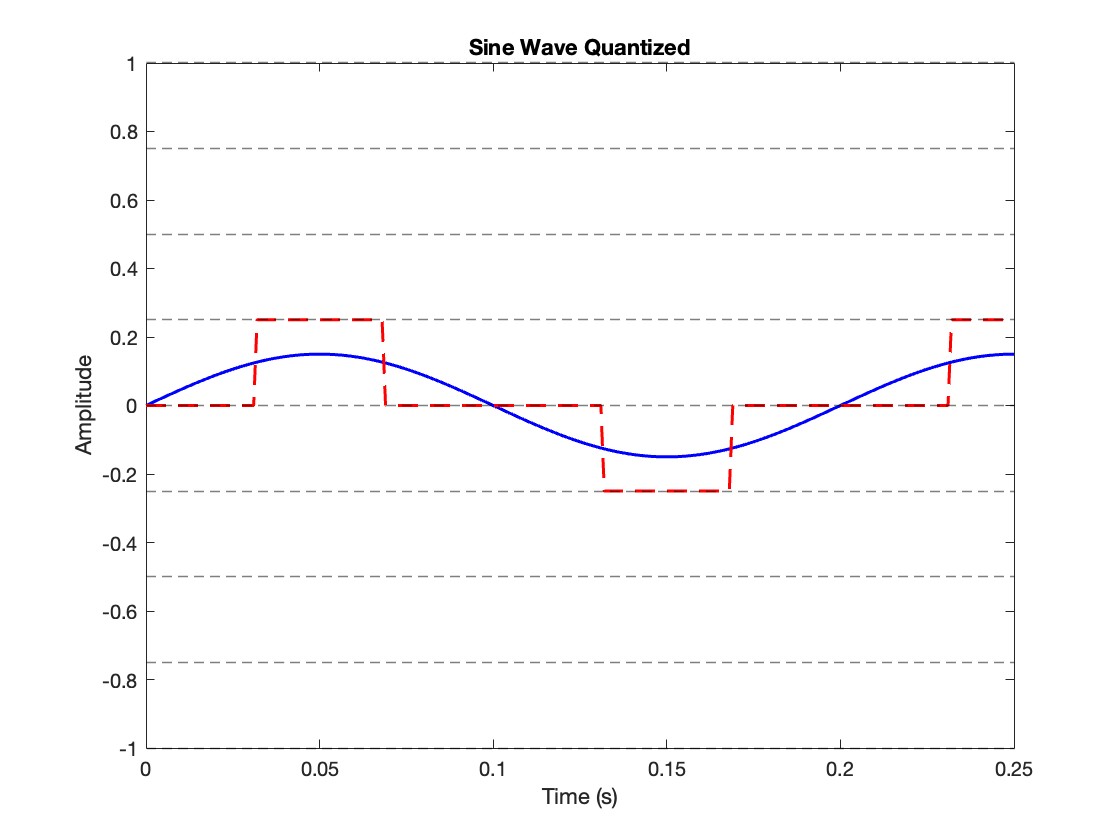} \\
{\small (a)} & {\small (b)}
\end{tabular}
\caption{\label{fig:6}%
(a) Sine wave with amplitude 1 (blue) overlaid by quantized sine wave to 3 bits (red). 
(b) Sine wave (amplitude 0.5, blue) with 3-bit quantization (red).}
\end{figure}

\indent For all of the tests, the quantization step size was held constant with $A$ set to 1, rather than being dependent on the maximum amplitude of the signal. This approach allowed the impact of the dither size to be investigated directly, for the loudness tests. If $A$ were dependent on the maximum amplitude of the input signal, it would be expected that the perceptual quality of the quantized signal would remain consistent despite changes in amplitude, provided the quantization bit depth remained constant. This is because smaller signals are less accurately preserved, as they fall within a narrower range of quantization levels than larger ones, leading to reduced fidelity in retaining the original signal shape (see Fig.~\ref{fig:6}).\ 

\indent For the pitch tests, it was anticipated that increasing the pitch would reduce the performance of the dithering technique. This decline in performance should not be attributed to aliasing effects, as the highest test frequency, 534.35 Hz, is well below the Nyquist rate of 22.05 kHz for these tests. However, if the tested audio sample is not a pure sine wave, harmonics could potentially exceed the Nyquist rate, leading to aliasing effects.\ 

\indent Additionally, psychoacoustic factors may explain this behavior. At higher frequencies, the perceived loudness of the signal is lower compared to lower-frequency signals, even when their physical amplitudes are identical. Since the dither noise is uniformly distributed across all frequencies, the perceived signal-to-noise ratio (SNR) decreases at higher frequencies. This reduction in perceived SNR negatively impacts the perceptual quality of the audio signal, contributing to the observed decrease in dithering performance with increasing pitch. However, given the small pitch range tested (C4 to C5), the difference in performance is expected to be minimal.\ 

\indent Two perceptual measures were used: VISQOL (Virtual Speech Quality Objective Listener) and STOI (Short-Time Objective Intelligibility), implemented using MATLAB's Audio Toolbox R2024b. VISQOL provides a general measure of audio quality, with scores ranging from 1 (worst) to 5 (best)~\cite{Hines2015}. STOI, while optimized for speech audio signals with scores ranging from 0 to 1, was also employed as a supplementary validation for the VISQOL results~\cite{Dong2020}.\ 

\begin{figure}[h]
\centering
\includegraphics[width=5in]{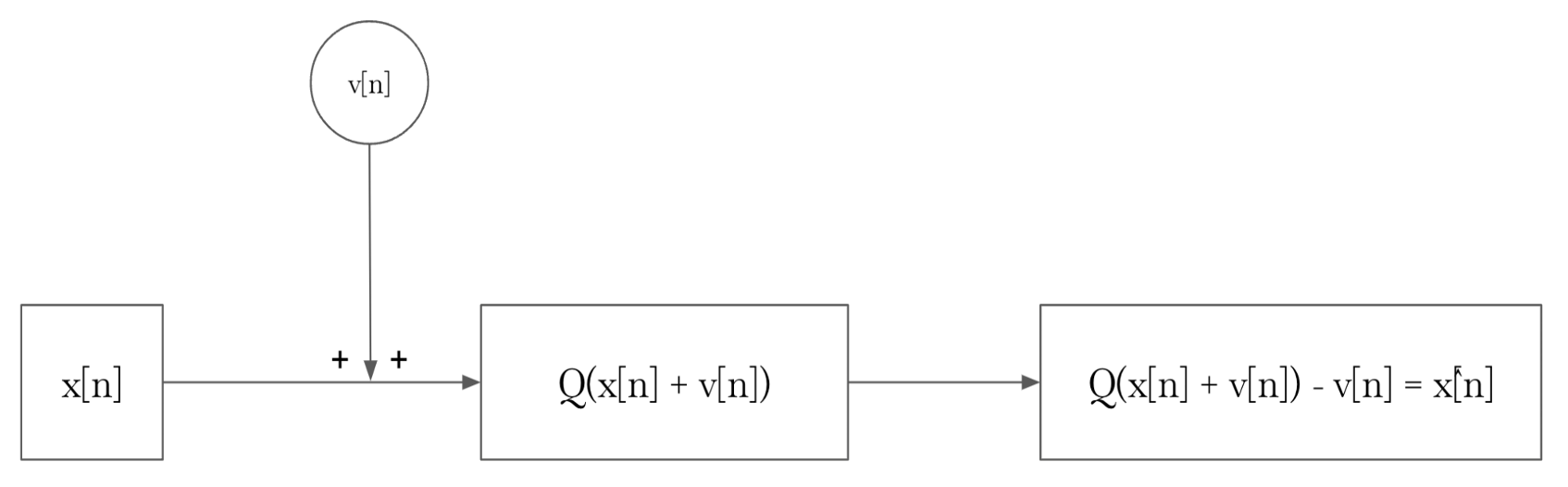}
\caption{\label{fig:7}%
Pipeline for Dither Quantizer.}
\end{figure}

\indent The tests included four different sets of conditions: TPDF, TPDF Shaping, TPDF Modified, and TPDF Modified Shaping. Before processing, all audio files were down-mixed to mono. Additionally, all input signals were normalized in MATLAB so that $A$ would be 1, with the exception of the audio files used in the loudness tests. For non-noise-shaping conditions, the processing pipeline involved adding a dither signal to the original audio signal, followed by quantization to 3 bits (see Fig.~\ref{fig:7}).\ 

\begin{figure}[h]
\centering
\includegraphics[width=5in]{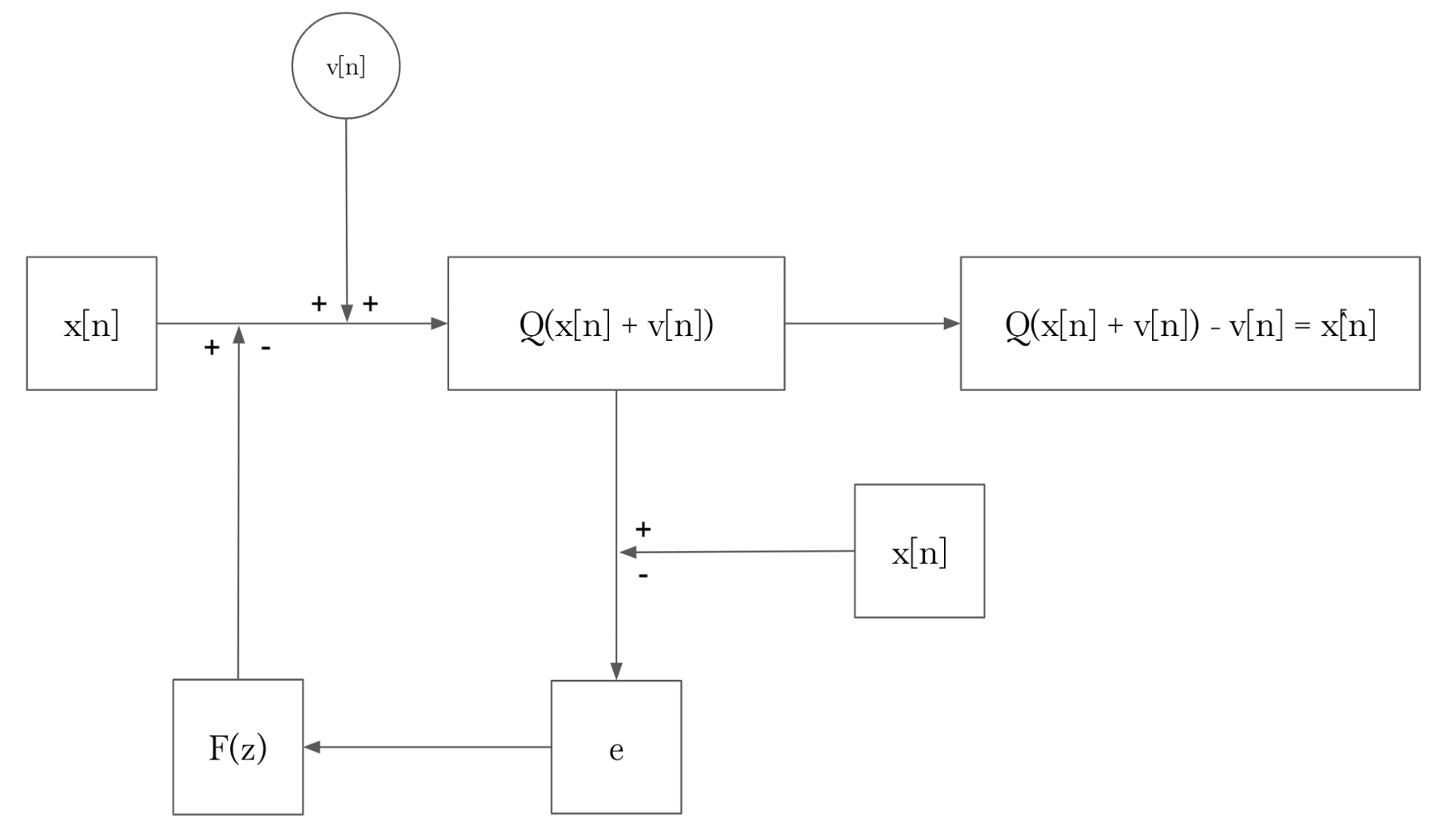}
\caption{\label{fig:8}%
Pipeline for Dither Quantizer with noise-shaping.}
\end{figure}

\indent The pipeline for the shaping conditions included a feedback loop. After quantization to 3 bits, the error signal was calculated, filtered by the inverse of the equal-loudness contour, and subtracted from the original signal over a number of n iterations (see Fig.~\ref{fig:8}). The filter was designed using a MATLAB function that generates a 512th-order FIR filter with a frequency response shaped to approximate the inverse of the equal-loudness contour. The function interpolates frequency-gain points based on the inverse of the equal-loudness contour, converts the gains from dB to a linear scale, and uses the \textit{fir2} function to create a filter that compensates for human hearing sensitivity across the audible spectrum.\

\indent To limit computational intensity, the project utilized only 100 iterations, and a set of 1000 alpha values was tested for all audio samples. The alpha values were uniformly sampled across the range from 0 to 1, encompassing both a null PDF condition and full TPDF dithering for each test. Additionally, RPDF dithering was tested as a baseline condition for comparison with the results of the four TPDF test conditions.\ 

\indent After processing the audio files, entropy, MSE, VISQOL, and STOI were calculated using either audio toolbox functions or custom functions based on the equations detailed in previous sections.\ 

\indent The compression and perceptual measures were all taken for both the non-subtrac-tive dithering output and the subtractive dithering output. Subtractive dither is a technique that removes the added noise after quantization by subtracting the exact noise signal from the output, minimizing distortion of the audio signal while preserving the benefits of dithering~\cite{Arar2023}. This should result in a significantly lower MSE with the SD condition compared to the NSD condition. With VISQOL and STOI being intrusive, $\hat{x}$ is the processed signal, while $x$ is the input signal exported from Logic Pro to 8 bits. Given that the reference signal is specific to each test condition, the scores provide a relative quality difference between the input signal and the test signal.\

\begin{figure}[h]
\centering
\begin{tabular}{cc}
\includegraphics[width=3in]{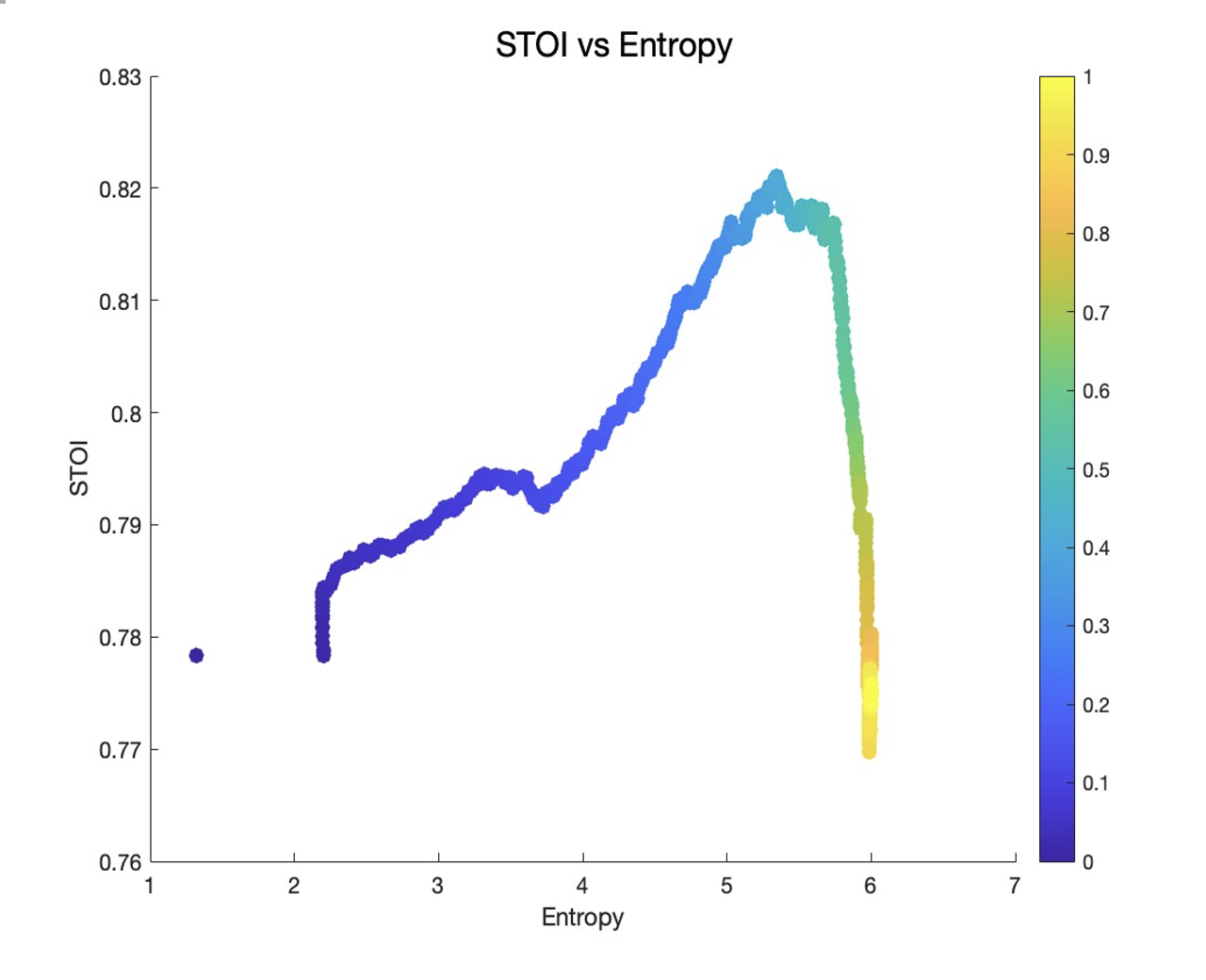} &
\includegraphics[width=3in]{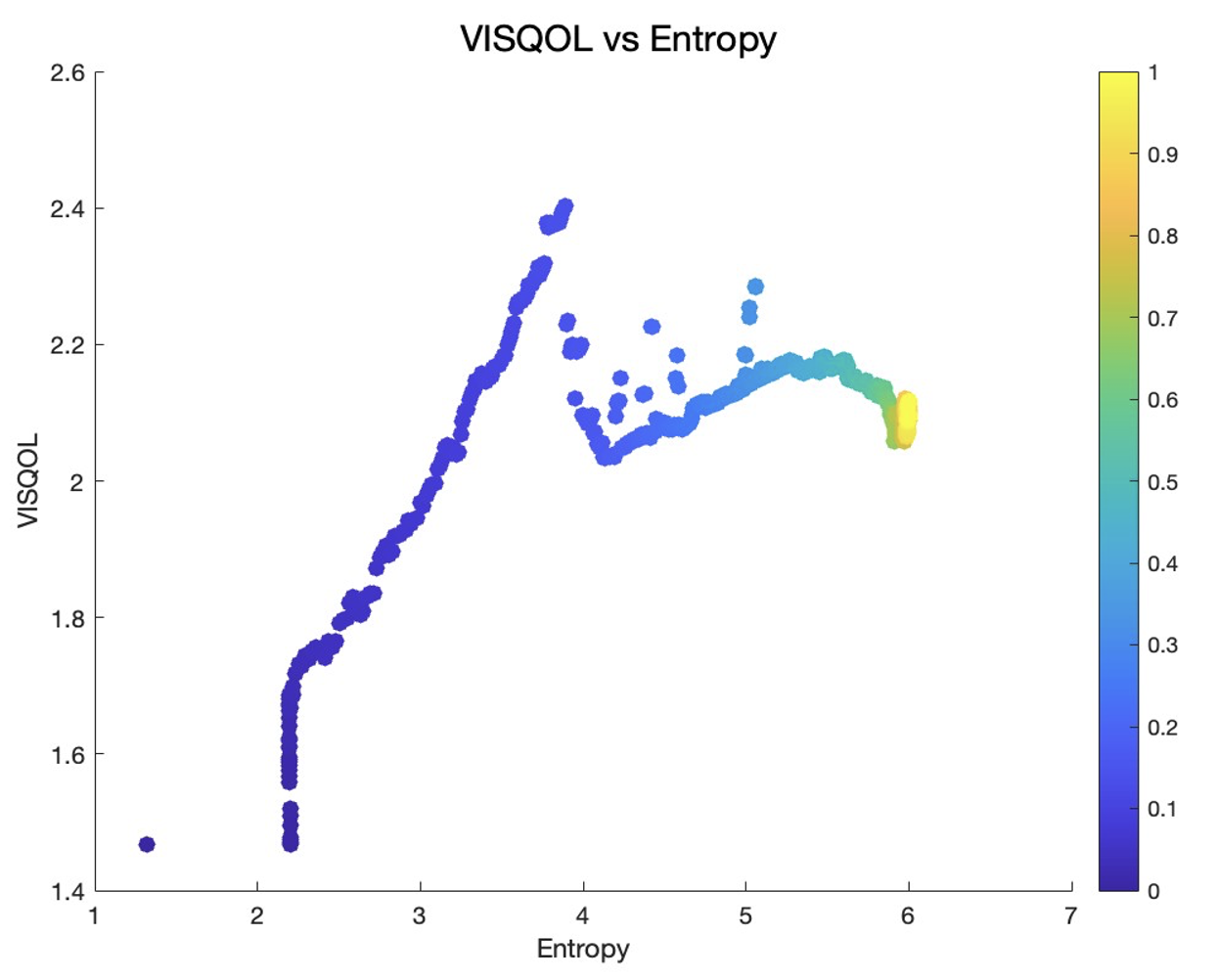} \\
{\small (a)} & {\small (b)}
\end{tabular}
\caption{\label{fig:9}%
(a) Piano chord Modified TPDF No Shaping STOI. (b) Piano chord Modified TPDF No Shaping VISQOL.}
\end{figure}

\Section{Results}
The optimal alpha value was determined by analyzing the entropy versus perceptual measure tests and identifying spikes or plateaus in the perceptual measure scores. Overall, there was a strong alignment between the STOI and VISQOL metrics in identifying the optimal alpha value for balancing trade-offs. However, occasional discrepancies were observed in the optimal alpha values determined by the two metrics, as demonstrated in the results for the piano chord sample (see Fig.~\ref{fig:9})\ 

\begin{figure}[h]
\begin{center}
\begin{tabular}{c}
\epsfig{width=4in,file=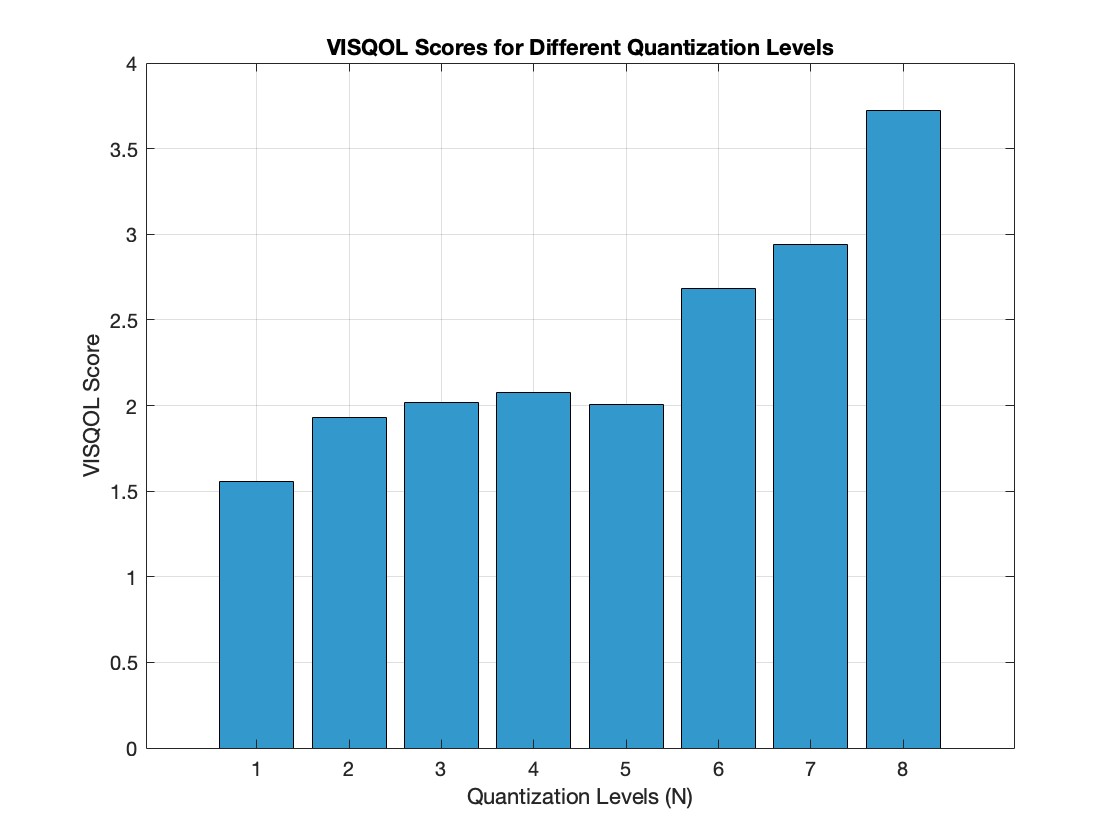} \\
\end{tabular}
\end{center}
\caption{\label{fig:10}%
VISQOL scores for the quantization of a 0 dB sine audio file without dithering at varying bit depths.}
\end{figure}

\indent Additionally, improvement in the score was quantified as the difference between the perceptual audio score for the no-dithering condition and the score at the optimal alpha value. For all tests, the final processed audio subject to analysis utilized subtractive dithering.

\indent For all test conditions, 3-bit quantization was applied, which resulted in low visqol scores just under 2. This outcome is expected given the coarse quantization the signal undergoes. To further investigate this behavior, higher-bit quantization was tested, producing higher VISQOL scores but with relatively small improvements. The trends observed with 3-bit quantization persisted with higher-bit quantization.\

\indent With higher-bit quantization, the quantization error is significantly reduced, resulting in higher audio quality and leaving less room for perceptual improvements, making it more difficult to show distinct results (see Fig.~\ref{fig:10}). This aligns with the nature of perceptual measures like VISQOL, which are intrusive and calculated by comparing the processed audio to the original samples [13].\

\indent For all test conditions, MSE and entropy were observed to increase with higher alpha values, while VISQOL scores demonstrated non-uniform behavior depending on the specific conditions. Across all tested loudness levels, the RPDF condition ranked lower than two or more of the TPDF dither distributions, with the null PDF consistently scoring the worst. RPDF consistently outperformed the Modified TPDF conditions. At higher loudness levels, both un-modified TPDF distribution conditions performed the best, whereas, at lower loudness levels, the distributions without noise-shaping showed superior performance to the distributions with noise-shaping (see Table~\ref{tab:1}).\ 

\begin{table}[h]
\begin{center}
\caption{\label{tab:1}%
(a) Results for VISQOL and Entropy at -10 dB (b) Results for VISQOL, Entropy, and Alpha at -20 dB.
}
{
\renewcommand{\baselinestretch}{1}\footnotesize
\begin{tabular}{|c|c|c|c|c|c|c|}
\cline{2-7}
\multicolumn{1}{c|}{~} &
\multicolumn{6}{c|}{(a) Results for VISQOL, Entropy, and Alpha at -10 dB}\\
\cline{2-7}
\multicolumn{1}{c|} ~ &
TPDF & Modified TPDF & TPDF Shaping & Modified TPDF Shaping & RPDF & NPDF\\
\hline
VISQOL &2.58 &2.08 &2.49 &1.97 &2.20 &1.78 \\
Entropy &4.28 &4.10 &4.38 &4.46 &4.14 &1.57 \\
\hline
\end{tabular}
\item
\begin{tabular}{|c|c|c|c|c|c|c|}
\cline{2-7}
\multicolumn{1}{c|}{~} &
\multicolumn{6}{c|}{(b) Results for VISQOL, Entropy, and Alpha at -20 dB.}\\
\cline{2-7}
\multicolumn{1}{c|} ~ &
TPDF & Modified TPDF & TPDF Shaping & Modified TPDF Shaping & RPDF & NPDF\\
\hline
VISQOL &1.89 &1.91 &1.74 &1.69 &1.86 &1.00 \\
Entropy &5.08 &2.87 &4.89 &5.15 &4.82 &1.53 \\
\hline
\end{tabular}}
\end{center}
\end{table}

\begin{figure}[!h]
\centering
\includegraphics[width=5in]{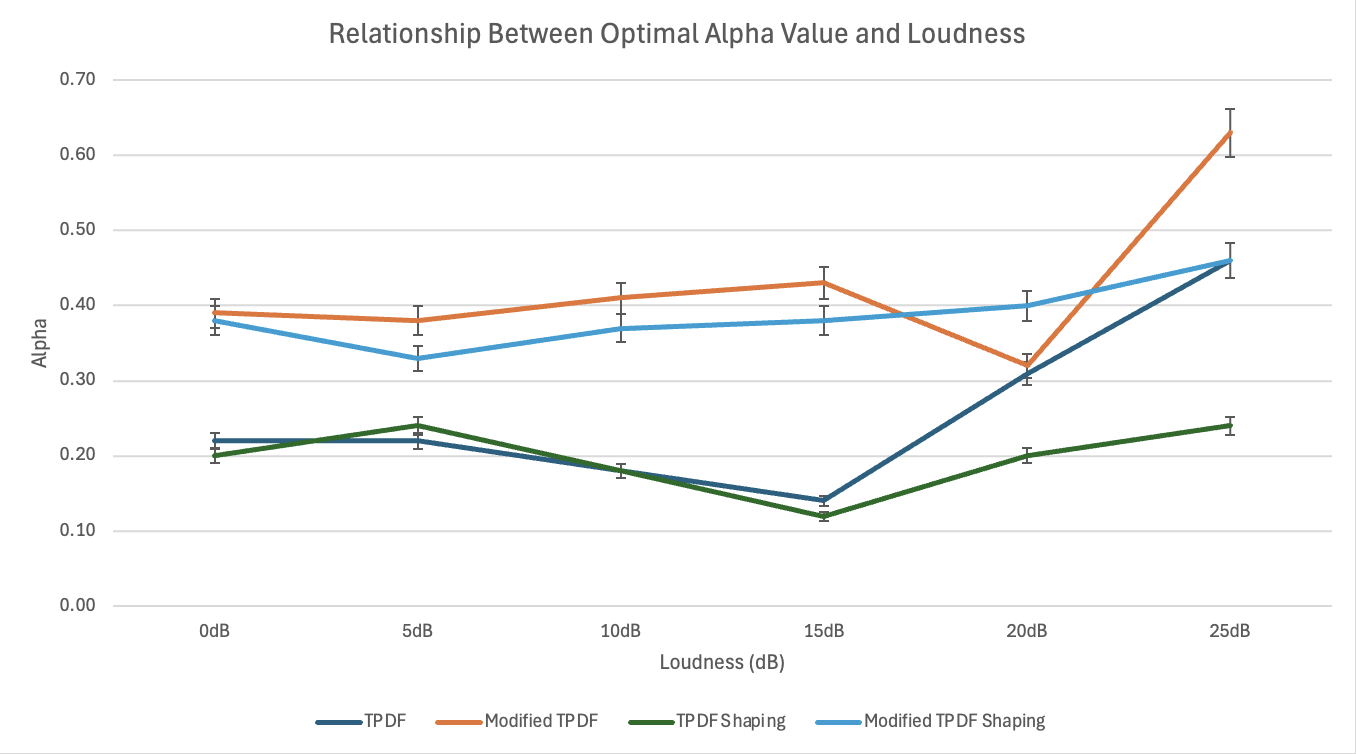}
\caption{\label{fig:11}%
Relationship between optimal alpha value and loudness for the four tested TPDF dither distributions.}
\end{figure}

\indent For each loudness level tested, the highest VISQOL scores varied across the different dither distributions, indicating that performance is not uniform and highlighting the relevance of loudness to the technique's effectiveness. Optimal alpha dithering generally demonstrated similar or better VISQOL performance compared to full dithering, with the added benefit of lower entropy values, underscoring the advantages of entropy-controlled dithering. The observed optimal alpha values remained consistent across tests, with a slight dip for regular TPDF conditions at -15 dB and a spike in non-shaping conditions at -25 dB (see Fig.~\ref{fig:11}).\

\begin{figure}[h]
\centering
\includegraphics[width=5in]{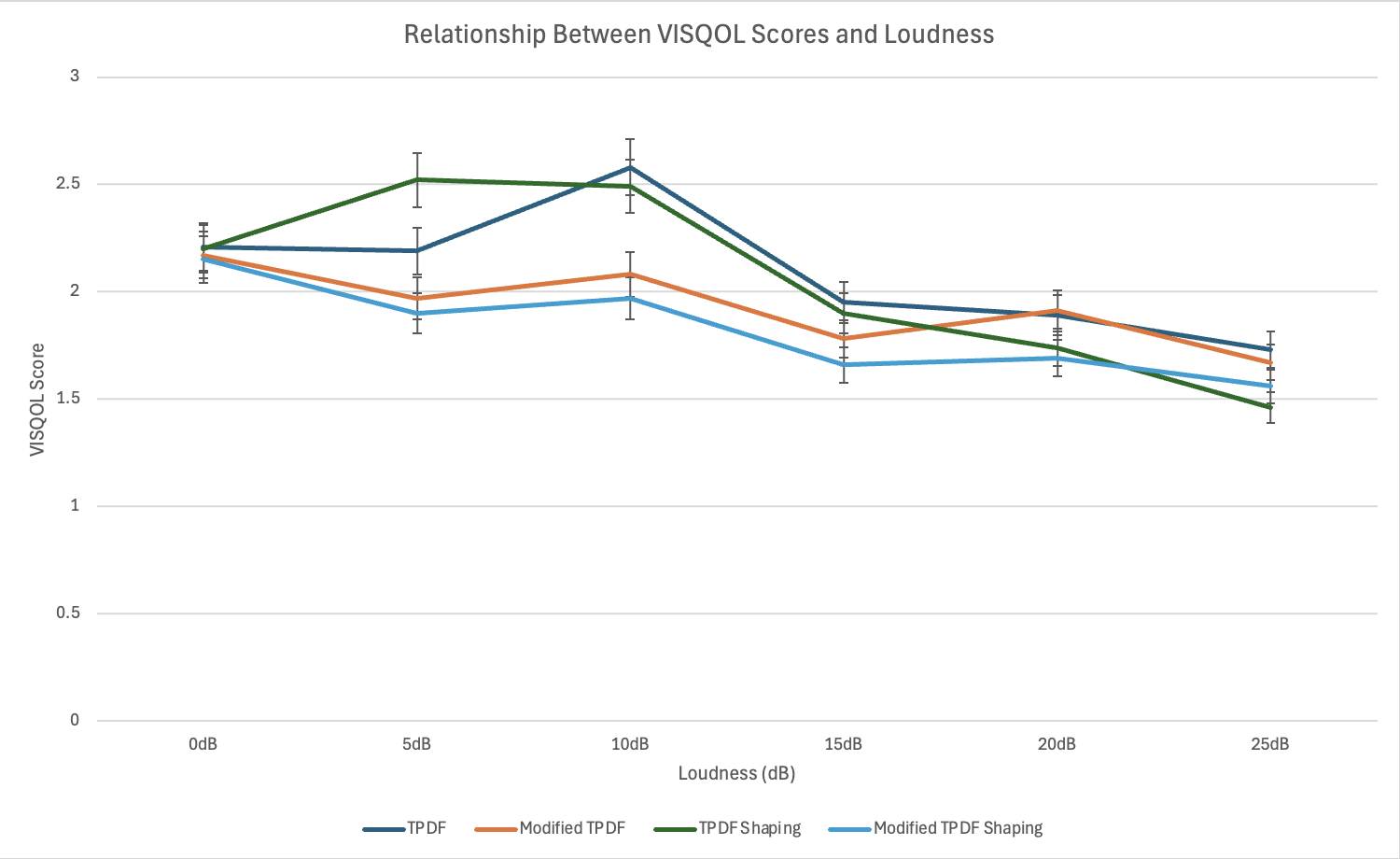}
\caption{\label{fig:12}%
Relationship between VISQOL scores and loudness for the four tested TPDF dither distributions.}
\end{figure}

\indent Entropy values for full dithering in no-shaping conditions were lower than those of shaping conditions at higher loudness levels, whereas, at lower loudness levels, entropy was higher for non-shaping conditions. Suggesting that noise shaping ability to effectively redistribute noise away from a uniform frequency spectrum is limited with lower amplitude signals. While no definitive trend was observed for the Modified TPDF conditions, a weak negative trend emerged for all TPDF distributions as loudness levels decreased, supporting the initial hypothesis (see Fig.~\ref{fig:12}).\ 

\begin{figure}[!h]
\centering
\begin{tabular}{cc}
\includegraphics[width=3in]{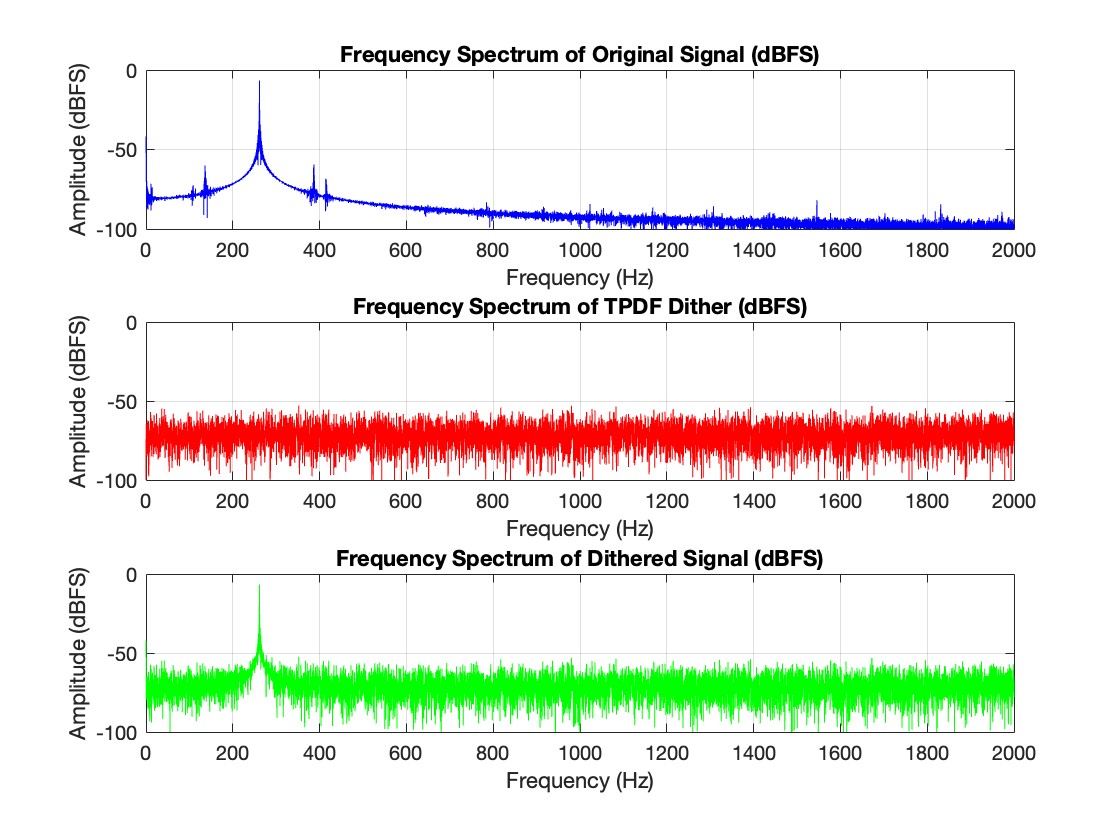} &
\includegraphics[width=3in]{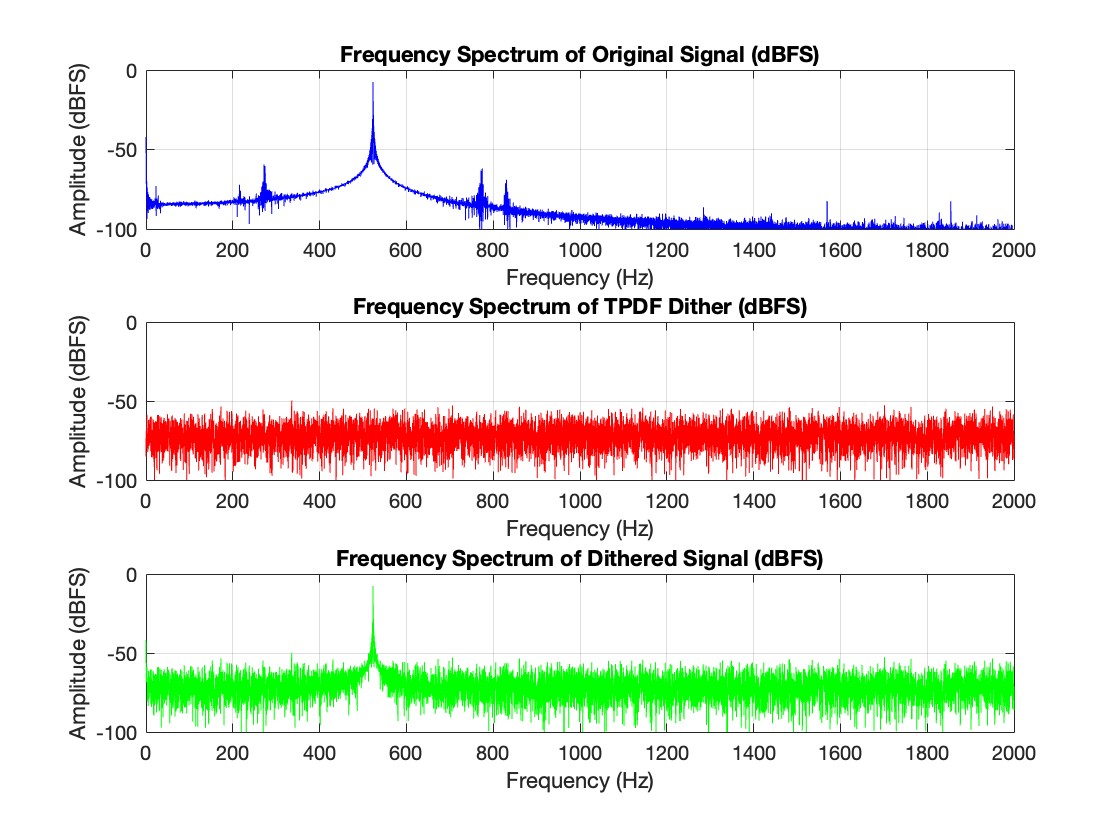} \\
{\small (a)} & {\small (b)}
\end{tabular}
\caption{\label{fig:13}%
(a) Frequency spectra for 3-bit quantized signals. Top: C4 spectrum with a strong fundamental at 261 Hz and visible harmonics. Middle: TPDF dither at optimal alpha value spectrum with a flat noise floor. Bottom: Dithered C4 spectrum with reduced harmonic distortion. (b) Top: C5 spectrum (523 Hz) with visible harmonics. Middle: TPDF dither spectrum. Bottom: Dithered C5 spectrum with reduced harmonic distortion.}
\end{figure}

\indent For tests conducted across various frequencies, analyzing the frequency spectrum of the signals provides a deeper understanding of observed trends. When this analysis is applied to the highest and closest tested frequencies, it reveals that the undithered C5 sinusoidal signal exhibits greater harmonic content than the C4 signal. However, applying a dither signal masks these harmonic frequencies (see Fig.~\ref{fig:13}).

\indent For these tests, the Modified TPDF conditions generally showed slightly lower VISQOL scores compared to the regular TPDF dithering conditions. However, no significant difference was observed between the shaping and non-shaping conditions. The scores remained consistent as the pitch increased. However, the improvement in VISQOL from applying NPDF to TPDF distributions at the optimal alpha decreased with increasing pitch, indicating that pitch influences the effectiveness of the dithering technique. A similar pattern was observed in this set of tests, where lower optimal alpha values were seen for both Modified and non-modified TPDFs (see Table~\ref{tab:2}).\ 

\begin{table}[h]
    \begin{center}
        \caption{\label{tab:2}%
        Mean optimal alpha values for TPDF dither distributions across varying frequency tests.}
        \renewcommand{\baselinestretch}{1}\footnotesize
        \begin{tabular}{|c|c|c|c|c|}
            \cline{2-5}
            \multicolumn{1}{c|} ~ &
            TPDF & Modified TPDF & TPDF Shaping & Modified TPDF Shaping\\
            \hline
            Alpha &0.26 &0.43 &0.20 &0.39\\
            \hline
        \end{tabular}
    \end{center}
\end{table}

\begin{figure}[!h]
\centering
\includegraphics[width=5in]{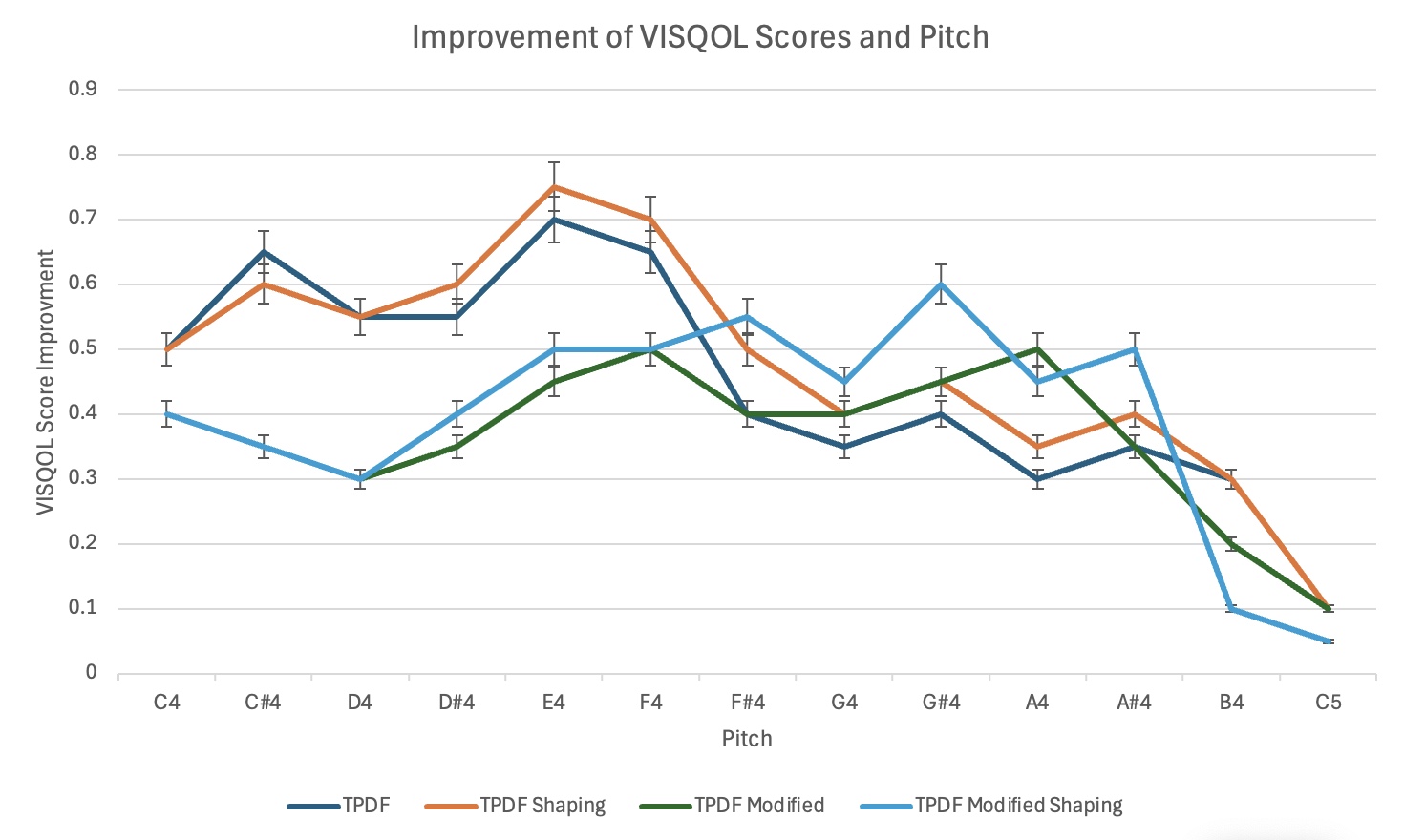}
\caption{\label{fig:14}%
Improvement of VISQOL score for varying pitches of sine audio sample.}
\end{figure}

\indent The data reveal an outlier for C5, which showed optimal conditions without dithering and no improvement when dithering was applied. The greatest improvement in the VISQOL score was observed for the lower tested frequencies, with a general decrease as the frequency increased (see Fig.~\ref{fig:14}).\ 

\indent In the third and fourth sections of the experiment, different instruments exhibited varying optimal alpha values and score improvements. However, the trends across the different conditions in terms of improving VISQOL scores were consistent in both sections (see Table~\ref{tab:3}).\ 

\begin{table}[!h]
\begin{center}
\caption{\label{tab:3}%
(a) Chord VISQOL Score Improvement (b) Rhythm VISQOL Score Improvement}
{
\renewcommand{\baselinestretch}{1}\footnotesize
\begin{tabular}{|c|c|c|c|c|}
\cline{2-5}
\multicolumn{1}{c|}{~} &
\multicolumn{4}{c|}{(a) Chord VISQOL Score Improvement}\\
\cline{2-5}
\multicolumn{1}{c|} ~ &
TPDF & Modified TPDF & TPDF Shaping & Modified TPDF Shaping\\
\hline
Piano &+1.01 &+0.39 &+0.95 &+0.53 \\
Trumpet &+0.83 &+0.51 &+0.63 &+0.40 \\
Sine &+0.58 &+0.62 &+0.51 &+0.41 \\
\hline
\end{tabular}
\item
\begin{tabular}{|c|c|c|c|c|}
\cline{2-5}
\multicolumn{1}{c|}{~} &
\multicolumn{4}{c|}{(b) Rhythm VISQOL Score Improvement}\\
\cline{2-5}
\multicolumn{1}{c|} ~ &
TPDF & Modified TPDF & TPDF Shaping & Modified TPDF Shaping\\
\hline
Piano &+0.79 &+0.75 &+0.61 &+0.43 \\
Trumpet &+0.22 &+0.25 &+0.14 &+0.05 \\
Bass &+0.21 &+0.26 &+0.13 &+0.04 \\
Drums &+0 &+0 &+0 &+0 \\
\hline
\end{tabular}}
\end{center}
\end{table}

\indent The results for these sections also demonstrated that dithering outperformed the no-dithering condition. While some tests indicated that full dithering yielded higher VISQOL scores, the entropy was significantly higher than that observed for the optimal alpha condition.\ 

\indent In the fourth section, where rhythm was introduced and multiple notes were played simultaneously, the no-noise-shaping condition outperformed the noise-shaping condition, although with a small sample size. Nevertheless, the Modified TPDF conditions showed slight improvements in VISQOL scores compared to the regular TPDF conditions at the optimal alpha value.\ 

\indent However, no definitive trend emerged regarding the performance of different distributions in the third and fourth tests, highlighting the unpredictability of using these distributions with the dithering technique.\ 

\indent For rhythmic sections, the introduction of rhythm generally resulted in no significant improvement across all conditions for a portion of the audio samples, suggesting that either the technique struggles with rhythm or perceptual quality testing has inherent limitations. Despite its overall poor performance, the Modified TPDF dithering condition outperformed the regular TPDF condition in speech audio tests for both shaping and non-shaping scenarios (see Table~\ref{tab:4}).\

\begin{table}[!h]
    \begin{center}
        \caption{\label{tab:4}%
        Speech STOI Score Improvement}
        \renewcommand{\baselinestretch}{1}\footnotesize
        \begin{tabular}{|c|c|c|c|c|}
            \cline{2-5}
            \multicolumn{1}{c|} ~ &
            TPDF & Modified TPDF & TPDF Shaping & Modified TPDF Shaping\\
            \hline
            Speech &+0.042 &+0.053 &+0.068 &+0.072\\
            \hline
        \end{tabular}
    \end{center}
\end{table}

\Section{Discussion}

\SubSection{Findings}
The results indicate that the optimal alpha values vary across different audio samples. While the dithering technique is consistently effective, its benefits are more pronounced under specific conditions. The TPDF dithering distributions achieved higher VISQOL scores compared to quantization without dithering in nearly all tests. Additionally, for most tests, all four TPDF distributions outperformed the RPDF distribution in terms of VISQOL, reflecting TPDF's superior noise-masking properties compared to RPDF.\

\indent In several tests within sections three and four, the addition of dither did not improve VISQOL scores. However, improvements were observed in STOI scores, highlighting a discrepancy between the two metrics. This raises the question of which metric should be prioritized. For consistency, VISQOL was selected as the primary evaluation metric.\ 

\indent In the first section of the test, a weak negative trend was observed in VISQOL scores as loudness levels decreased. While a slight decline in VISQOL performance was noted with decreasing loudness, the data does not strongly support the hypothesis. The predicted trend suggested a more substantial performance decrease; the lack of this substance performance decrease may be attributed to how VISQOL calculates its score relative to the input reference signal. Although the perceptual quality of the reference signal diminishes as amplitude decreases, the perceptual quality of the processed signal relative to its respective input signal does not decline to the same extent. That said, alternative explanations for the slight negative trend in performance may exist, warranting further investigation.\ 

\indent Additionally, in the second section, the results indicated that the technique produced similar VISQOL scores regardless of pitch. However, higher-frequency audio samples exhibited smaller improvements in VISQOL scores compared to NPDF conditions. This behavior is not simply due to less room for improvement but is likely influenced by the greater sensitivity of higher-frequency samples to the addition of noise or dithering, which impacts the perceptual SNR.\ 

\indent The hypothesis that perceptual SNR alone accounts for the behavior does not fully explain the significant drop in performance observed between C4 and C5, suggesting additional factors are involved. Spectral analysis of the lowest and highest tested frequencies revealed the presence of harmonics in the audio samples, indicating that the test signals were not pure sine waves. The higher-frequency samples exhibited greater harmonic content, likely influenced by system non-linearities or quantization artifacts. This harmonic content interacts with the dither noise floor, resulting in increased harmonic distortion, which offers a more plausible explanation for the observed performance decline  (see Fig. 13).\

\indent The findings underscore the trade-off between entropy and perceptual quality, while also demonstrating that noise shaping effectively enhances the dithering technique. The proposed method is relevant across diverse audio contexts, showcasing its broad applicability.\

\SubSection{Limitations}
This project encountered computational limitations, particularly in the noise-shaping feedback loop, which was constrained to only 100 iterations. This represents an area for future exploration, as increasing the number of iterations could yield more precise results and better simulate realistic audio compression scenarios.\ 

\indent Additionally, the tests conducted in this project involved truncating audio files to 3 bits—significantly lower than the resolutions typically used in real-world audio compression. While this approach effectively highlighted the trade-off between compressibility and perceptual quality scores, the findings assume that similar trends would apply to higher-resolution audio files.\

\indent There were also slight limitations in the perceptual quality tests. Some audio samples appeared to challenge the evaluation methods, suggesting potential shortcomings. These issues could be addressed by incorporating additional perceptual evaluation frameworks, such as the PEAQ, to ensure more robust and comprehensive assessments.\ 

\SubSection{Applications}
The increase in the compressibility of high-quality audio files offers significant benefits across various audio applications beyond music. One such area is machine learning, where substantial amounts of audio data must be processed efficiently. By maximizing the storage and transfer of audio data, this advancement enhances the feasibility of handling large datasets required for training and inference in machine-learning applications~\cite{Saseendran2019}.\ 

\indent Moreover, this development is particularly valuable in emerging technologies such as Mixed Reality (MR). High-quality media is crucial for MR applications, where efficient data compression minimizes artifacts and ensures a seamless and immersive user experience. This optimization is essential for balancing the demands of speed and storage in such applications~\cite{Son2020}.\ 

\indent Communication technologies also stand to benefit from improved audio compressibility. Enhanced telecommunication speeds, through platforms like Zoom and Skype, can improve accessibility and connectivity. This is especially impactful for underserved communities, where access to quality communication tools is critical for education and community-building initiatives~\cite{Graves2021}.\ 

\indent In addition, medical applications, such as hearing aids and assistive listening devices, could leverage these advancements. Reduced storage requirements for higher-quality audio translate to lower latency, reduced delays in sound amplification, and less bandwidth needed for audio transmission. These improvements enhance the reliability of wireless connections and the overall user experience~\cite{Balling2020}.\ 

\begin{figure}[h]
\centering
\includegraphics[width=4in]{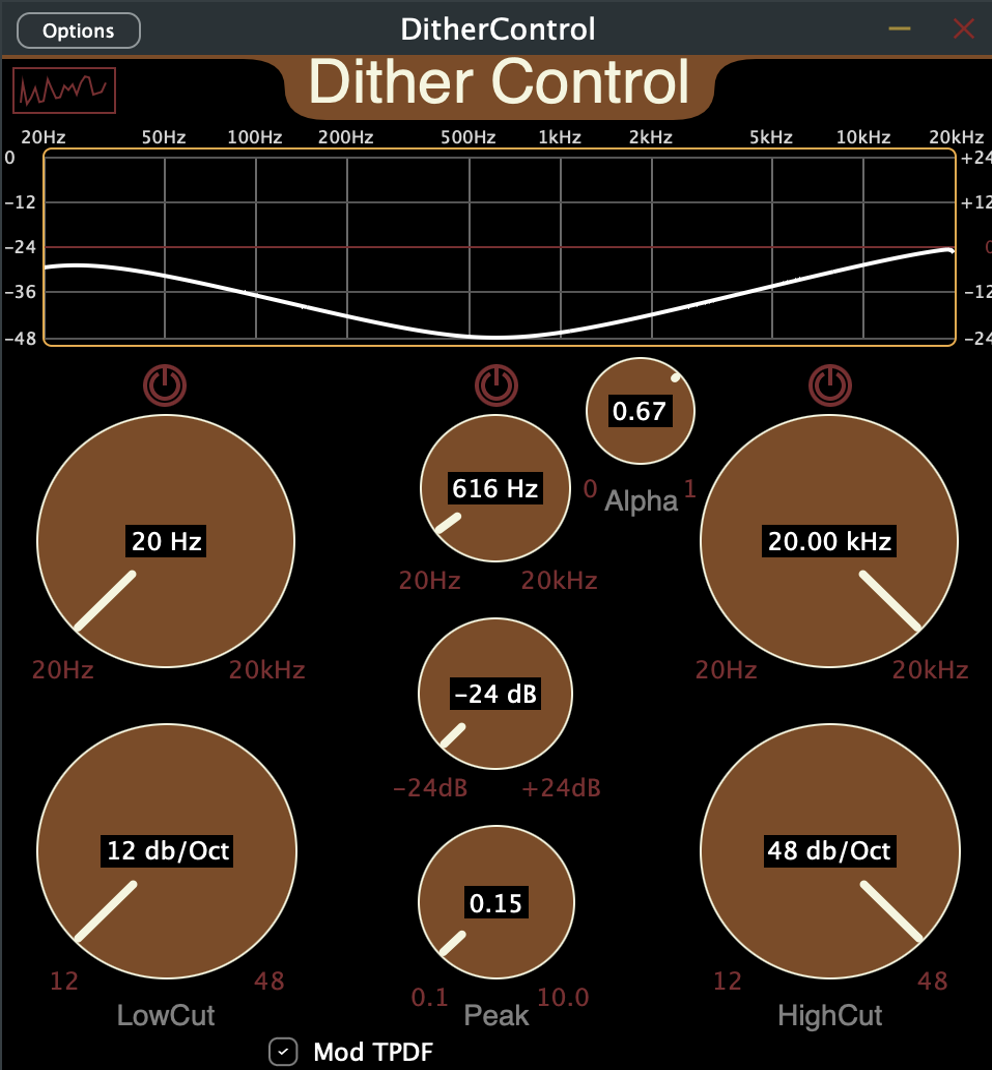}
\caption{\label{fig:15}%
Entropy-Controlled Dithering Audio Plugin.}
\end{figure}

\SubSection{Dither Control Plugin}
What’s next for this project is making these findings accessible and applicable in audio use. This could be done directly through its implementation as a user-friendly plugin compatible with a wide range of Digital Audio Workstations (see Fig.~\ref{fig:15})\ 

\indent This plugin, developed in C++ using the JUCE framework, provides accessible control over a dither design technique informed by the findings of this project. It features controls for designing a noise-shaping filter using a parametric EQ and an alpha parameter to adjust the amplitude of the dither signal. \ 

\indent The goal is to give audio engineers and producers greater flexibility in the dithering process by enabling the application of distortion dithering to individual tracks, rather than relying on a standardized dither for the entire project. With further tests of this entropy-controlled dithering technique across a wider range of audio samples, additional presets can be developed, increasing the applicability of this useful plugin. \ 

\Section{Conclusion}
This paper demonstrates the effectiveness of an entropy-compression-controlled dithering technique for audio compression, leveraging two distinct dither distributions. The study highlights the rationale for applying this approach across various audio scenarios, including changes in rhythm, pitch, and loudness, and identifies conditions where specific distributions are more suitable. The findings reveal that pitch and loudness significantly influence the technique's performance: higher pitch diminishes its effectiveness, while lower loudness also reduces its impact. However, when analyzing alternate rhythms and multiple frequencies, careful consideration of each audio signal's unique characteristics is essential.\ 

\indent Future research could further investigate entropy-controlled dithering by examining its interaction with nonuniform quantization techniques, such as Lloyd-Max Quantization or µ-law companding. Tailoring an optimal dither distribution for these nonuniform quantizers could enhance their performance, paving the way for the development of a novel audio compression algorithm. \

\Section{Author Declarations}
The authors declare no conflicts of interest.

\Section{References}
\bibliographystyle{IEEEbib}
\bibliography{refs}

\begin{thebibliography}{10}

\bibitem{Ahmed2018}
R.~Ahmed, M.~Sazzadul, and J.~Uddin,
\newblock ``Optimizing apple lossless audio codec algorithm using nvidia cuda architecture,''
\newblock {\em International Journal of Electrical and Computer Engineering}, vol. 8, no. 1, pp. 70--75, 2018,
\newblock [Online]. Available: https://doi.org/10.11591/ijece.v8i1.pp70-75.

\bibitem{Ng1998}
W.~K. Ng, S.~Choi, and C.~V. Ravishankar,
\newblock ``Lossless and lossy data compression,'' 1998.

\bibitem{Kasher2024}
M.~Kasher, M.~Tinston, and P.~Spasojevic,
\newblock ``Distortion-controlled dithering with reduced recompression rate,'' 2024,
\newblock [Online]. Available: https://doi.org/10.48550/arXiv.2402.16447.

\bibitem{Floros2006}
A.~Floros and M.~Avlonitis,
\newblock ``Advances on calculating effective dither for audio signals,''
\newblock in {\em Proc. of the 10th WSEAS Int. Conf. on Systems}, Vouliagmeni, Athens, Greece, 2006, pp. 614--618,
\newblock [Online]. Available: https://api.semanticscholar.org/CorpusID:39647275.

\bibitem{Borsky2016}
M.~Borsky, P.~Mizera, P.~Pollak, and J.~Nouza,
\newblock ``Dithering techniques in automatic recognition of speech corrupted by mp3 compression: Analysis, solutions, and experiments,''
\newblock {\em Speech Communication}, vol. 86, pp. 40--50, 2016,
\newblock [Online]. Available: https://doi.org/10.1016/j.specom.2016.11.007.

\bibitem{Vanderkooy1987}
J.~Vanderkooy and S.~P. Lipshitz,
\newblock ``Dither in digital audio,'' 1987,
\newblock [Online]. Available: https://aes2.org/publications/elibrary-page/?id=5173.

\bibitem{Gray1998}
R.~M. Gray and D.~L. Neuhoff,
\newblock ``Quantization,''
\newblock {\em IEEE Transactions on Information Theory}, vol. 44, no. 6, pp. 2325--2383, 1998,
\newblock [Online]. Available: https://doi.org/10.1109/18.720541.

\bibitem{Menkman2011}
R.~Menkman,
\newblock {\em The glitch moment(um)},
\newblock Institute of Network Cultures, 2011,
\newblock [Online]. Available: https://doi.org/ISBN978-90-816021-6-7.

\bibitem{Herre1999}
J.~Herre,
\newblock ``Temporal noise shaping, quantization, and coding methods in perceptual audio coding: A tutorial introduction,''
\newblock in {\em AES 17th Int. Conf. on High Quality Audio Coding}, 1999,
\newblock [Online]. Available: https://aes2.org/publications/elibrary-page/?id=8057.

\bibitem{Kulkarni2014}
S.~R. Kulkarni,
\newblock ``Ele 201: Information signals - course notes, chapter 8: Information, entropy, and coding,'' 2014.

\bibitem{Hines2015}
A.~Hines, J.~Skoglund, A.~Kokaram, and N.~Harte,
\newblock ``Visqol: an objective speech quality model,''
\newblock {\em EURASIP Journal on Audio, Speech, and Music Processing}, vol. 2015, 2015.

\bibitem{Dong2020}
X.~Dong and D.~S. Williamson,
\newblock ``Towards real-world objective speech quality and intelligibility assessment using speech-enhancement residuals and convolutional long short-term memory networks,''
\newblock {\em The Journal of the Acoustical Society of America}, vol. 148, no. 5, pp. 3348, 2020.

\bibitem{Arar2023}
S.~Arar,
\newblock ``Reducing quantization distortion via subtractive and non-subtractive dithering,'' 2023,
\newblock [Online]. Available: https://www.allaboutcircuits.com/technical-articles/reducing-quantization-distortion-via-subtractive-and-non-subtractive-dithering [Accessed: Nov. 7, 2024].

\bibitem{Saseendran2019}
A.~T. Saseendran, L.~Setia, V.~Chhabria, and A.~B. Roy,
\newblock ``Impact of noise in the dataset on machine learning algorithms,'' 2019,
\newblock [Online]. Available: https://doi.org/10.13140/RG.2.2.25669.91369.

\bibitem{Son2020}
J.~Son, W.~Morgenstern, P.~Eisert, S.~Gül, A.~Hilsmann, T.~Schierl, G.~S. Bhullar, T.~Ebner, T.~Buchholz, G.~Hege, S.~Bliedung, and C.~Hellge,
\newblock ``Split rendering for mixed reality: Interactive volumetric video in action,''
\newblock in {\em Fraunhofer HHI}, 2020,
\newblock [Online]. Available: https://doi.org/10.1145/3415256.3421491.

\bibitem{Graves2021}
J.~M. Graves, D.~A. Abshire, S.~Amiri, and J.~L. Mackelprang,
\newblock ``Disparities in technology and broadband internet access across rurality: Implications for health and education,''
\newblock {\em Family \& Community Health}, vol. 44, no. 4, pp. 257--265, 2021,
\newblock [Online]. Available: https://doi.org/10.1097/FCH.0000000000000306.

\bibitem{Balling2020}
L.~W. Balling, O.~Townend, G.~Stiefenhofer, and W.~Switalski,
\newblock ``Reducing hearing aid delay for optimal sound quality: A new paradigm in processing, behind the ear, in the ear, speech in noise?,'' 2020,
\newblock [Online]. Available: https://hearingreview.com/hearing-products/hearing-aids/bte/reducing-hearing-aid-delay-for-optimal-sound-quality-a-new-paradigm-in-processing.

\end{thebibliography}

\end{document}